From secretome analysis to immunology: Chitosan induces major alterations in the activation of dendritic cells via a TLR4-dependent mechanism

Christian Villiers†*, Mireille Chevallet‡,§*, Hélène Diemer¶, Rachel Couderc †, Heidi Freitas†, Alain Van Dorsselaer¶, Patrice N. Marche †, Thierry Rabilloud‡,§

†: INSERM U823 Analytical Immunology of Chronic Pathologies, Institut Albert Bonniot, BP170, 38042 Grenoble, France, and Université Joseph Fourier, France

‡ CEA-DSV/iRTSV/LBBSI, Biophysique et Biochimie des Systèmes Intégrés, CEA-Grenoble, 17 rue des martyrs, F-38054 GRENOBLE CEDEX 9, France

§ CNRS UMR 5092, Biophysique et Biochimie des Systèmes Intégrés, CEA-Grenoble, 17 rue des martyrs, F-38054 GRENOBLE CEDEX 9, France

¶: CNRS UMR7178, Institut Pluridisciplinaire Hubert Curien, ECPM, 25 rue Becquerel, F-67087 STRASBOURG Cedex2, France

* these authors contributed equally to this work

Correspondence :
Thierry Rabilloud, iRTSV/LBBSI
CEA-Grenoble, 17 rue des martyrs,
F-38054 GRENOBLE CEDEX 9
Tel (33)-4-38-78-32-12
Fax (33)-4-38-78-44-99
e-mail: Thierry.Rabilloud@ cea.fr

Abstract

Dendritic cells are known to be activated by a wide range of microbial products, leading to cytokine production, and increased levels of membrane markers such as MHC Class II molecules. Such activated dendritic cells possess the capacity to activate naïve T cells. We demonstrate here that immature dendritic cells secrete both the YM1 lectin and lipocalin2. By testing the ligands of these two poteins, respectively chitosan and siderophores, we also demonstrate that chitosan, a degradation product of various fungal and protozoal cell walls, induce an activation of dendritic cells at the membrane level, as shown by the upregulation of membrane proteins such as class II molecules, CD80 and CD86, via a TLR4-dependent mechanism, but is not able to induce cytokine production. This leads to the production of activated dendritic cells unable to stimulate T cells. However, costimulation with other microbial products overcomes this partial activation and restore the capacity of these activated dendritic cells to stimulate T cells. In addition, successive stimulation with chitosan and then by LPS induced a dose-dependent change in the cytokinic IL-12/IL-10 balance produced by the dendritic cells.

# 1. Introduction

Appropriate sensing of pathogens is a critical step in launching an adequate immune response. The immune system has evolved to elaborate a complex system of pathogen surveillance, in which specialized molecules (called pathogen recognition receptors) and specialized cell types (e.g. dendritic cells) play important roles. Dendritic cells (DCs) are very potent antigen-presenting cells. Upon stimulation by danger signals [1], including signals produced by external pathogens but also signals derived from abnormal cells [2], DC undergo activation/maturation to an active antigen-presenting phenotype. This phenotype includes transfer to the cell surface of antigen-laden MHC class II molecules (also called signal 1), up-regulation of co-stimulatory membrane proteins, such as CD80 and CD86 (also called signal 2), as well as secretion of pro-inflammatory cytokines, such as TNFa, IL6 or IL12 (also called signal 3). Recent evidence has shown that all three signals are needed to induce T lymphocyte activation [3].

Most danger signals are sensed through receptors belonging to the Toll-like receptors (TLR) family [4]. These receptors are able to bind a wide range of ligands, and most of the known ligands are bacterial products (e.g. glycolipids, lipopolysaccharides) or viral ones (e.g. double stranded RNA. However, a few endogenous -TLR ligands, such as anionic polysaccharides [5-7] are also known, and may explain part of the behavior of dendritic cells toward endogenous activation signals [2]. However, little is known about the danger signals coming from eukaryotic pathogens (e.g. fungi, protozoa etc…) that are sensed by dendritic cells. TLR have been implicated in the sensing of fungal polysaccharides by monocytes [8] [9] and dendritic cells [10]. However, there are only very few examples in which the TLR involved in fungal polysaccharide recognition has been characterized [11]. Moreover, there seems to be several pathways for polysaccharides sensing, leading to different DC phenotypes [12, 13].

Moreover, sensing of danger signals sometimes involves other molecules than the ligand and the TLR, as shown by the complex sensing of LPS, in which protein mediators such as CD14 [14] and MD2 [15] are also implicated.

To gain insights into danger sensing by DC, we decided to perform proteomics experiments on proteins secreted by immature dendritic cells, looking for secreted proteins binding putative danger signals. We found proteins able to bind bacterial siderophores and chitosans, and further characterized the effects of these ligands on dendritic cells.

## 2. Material and methods

2.1. Cell cultures for proteomics experiments

Murine dendritic cells were produced from bone marrow progenitors. C57BL/6 mice were purchased from Charles River (l'arbresle, France). DC were generated from bone marrow as described [16], [17]. Briefly, bone marrow cells were isolated by flushing from the femurs. Erythrocytes and GR1 positives cells were removed by magnetic cell sorting. The remaining negatively sorted
cells were resuspended at $5 \times 10^5$ cells/ml in complete Iscove's modified Dubelcco's medium (IMDM) supplemented with 1% of GM-CSF-transfected J558cell line supernatant (This cell line was a generous gift of Dr David Gray, London), 40ng/ml of mouse recombinant FLT-3L and 5ng/ml of mouse recombinant Il6. At Day 3, the cellular supernatant was removed and the cells resuspended in the same conditions. From day 6 to day 11, IL6 was removed and FLT3-L was reduced to 20ng/ml. At day 11, the bone marrow cells are differentiated in DC and ready for the various experiments. ~~In some control experiments, dendritic cells were produced without FLT3-L, resulting in a much lower yield of cells~~. The resulting cells are essentially CD8 positive, and their cytokinic profile upon LPS activation is [IL6/IL12/TNFα]-high, [IL10/IFNγ]-low [18]. In some control experiments, dendritic cells were cultured without FLT3-L, leading to a much lower yield of cells. However, these cells had similar membrane markers and the same capacity to secrete cytokines. The cells were washed (three times in PBS, and then three times in serum-free DMEM ) and transferred to serum- and phenol red-free medium(1 million cells /ml. When needed the dendritic cells were activated by the addition of 1 μg of E. coli lipopolysaccharide or 40μg of chitosan per ml of culture medium for 24 hours. The cell supernates were then aspirated, collected ,and centrifuged at 1000g for 5 minutes (4°C) to pellet detached cells and large debris. The supernatant was collected and then centrifuged for 1 hour at 100,000g (4°C) to pellet smaller debris and vesicles. The final supernatant was saved and kept frozen (-80°C) until use.

For protein concentration [18] 20 ml of conditioned medium were first cooled on ice in a high speed centrifuge tube. Sodium lauroyl sarcosinate was then added to a final concentration of 0.1%. After mixing, TCA was added to a final 7.5% concentration, and the solution was precipitated on ice for 2 hours. The mixed protein-detergent precipitate was collected by centrifugation (10,000g, 10 minutes at 4°C). The supernatant was carefully removed, 2 ml of tetrahydrofuran (precooled in ice) were added to the pellet and vortexing was carried out until the pellet unstuck from the bottom of the tube, and dissolved almost completely. Centrifugation was carried out as described above. The supernatant was removed, and the nearly invisible pellet was washed again with 2ml of THF. Finally, the pellet was redissolved in 0.4 ml extraction solution (7M urea, 2M thiourea, 4% CHAPS, 0.4% carrier ampholytes (Pharmalytes 3-10) and 5mM Tris-carboxyethyl phosphine) with the help of a sonicator bath (30 minutes extraction).

All experiments were performed at least three times arising from different culture batches.

Chitin (from crab shells), chitosan (>85% deacetylated) and oligochitosan lactate were purchased from Sigma-Adrich. The E. coli siderophore was prepared from a culture supernatant as described previously [19], starting from a *fur*- coli strain to maximize siderophore production. To take into account possible contaminants of the siderophore preparation, a control preparation was made from a wild type coli strain with very low siderophore production.

2.2. Flow cytometry

The following mAbs were purchased from Pharmingen (San Diego, CA): anti-CD11b (M1/70, FITC-conjugated), anti-CD11c (HL3, APC-conjugated), anti-CD80 (16-10A1, PE-conjugated), anti-CD86 (GL1, PE-conjugated), anti-I-Ab (1F6-120.1, FITC-conjugated), Dead cells were excluded from the analysis by the absence of 7-AAD (7-Amino-Actinomycin D) staining. The analyses were performed on a FACS Calibur and on a FACSAria (BD Bioscience) and the analysis with FCS Express V3 (De Novo Software, Thornhill, Ontario, Canada).

2.3. Electrophoresis

IEF

Home made 160mm long 4-8 or 3-10.5 linear pH gradient gels were cast according to published procedures [20]. Four mm-wide strips were cut, and rehydrated overnight with the sample, diluted in a final volume of 0.6 ml of rehydration solution (7M urea, 2M thiourea, 4% CHAPS, 0.4% carrier ampholytes (Pharmalytes 3-10) and 100mM dithiodiethanol [21, 22]).
The strips were then placed in a multiphor plate, and IEF was carried out with the following electrical parameters

100V for 1 hour, then 300V for 3 hours, then 1000V for 1 hour, then 3400 V up to 60-70 kVh.

After IEF, the gels were equilibrated for 20 minutes in Tris 125mM, HCl 100mM, SDS 2.5%, glycerol 30% and urea 6M. They were then transferred on top of the SDS gels and sealed in place with 1% agarose dissolved in Tris 125mM, HCl 100mM, SDS 0.4% and 0.005% (w/v) bromophenol blue.

SDS electrophoresis and protein detection

10%T gels (160x200x1.5 mm) were used for protein separation. The gel buffer system is the classical Laemmli buffer pH 8.8, used at a ionic strength of 0.1 instead of the classical 0.0625. The electrode buffer is Tris 50mM, glycine 200mM, SDS 0.1%.

The gels were run at 25V for 1hour, then 100V until the dye front has reached the bottom of the gel. Detection was carried out by ammoniacal silver staining [23]. The spots of interest were excised by a scalpel blade and transferred to a 96 well microtitration plate. Destaining of the spots was carried out by the ferricyanide-thiosulfate method [24] on the same day than silver staining to improve sequence coverage in the mass spectrometry analysis [25]

2.4. Mass spectrometry

In-gel digestion :
Excised gel slice rinsing was performed by the Massprep (Micromass, Manchester, UK) as described previously [26]. Gel pieces were completely dried with a Speed Vac before digestion. The dried gel volume was evaluated and three volumes trypsin (Promega, Madison, US) 12.5ng/µl freshly diluted in 25mM $NH_4HCO_3$, were added. The digestion was performed at 35°C overnight. Then, the gel pieces were centrifuged for 5 min in a Speed Vac and 5µl of 29 100% H2O/ 69% acetonitrile/ 1% HCOOH were added to extract peptides. The mixture was sonicated for 5 min and centrifuged for 5 min. The supernatant was recovered and the procedure was repeated once.

MALDI-MS

MALDI-TOF mass measurements were carried out on Ultraflex™ TOF/TOF (Bruker DaltonikGmbH, Bremen, Germany). This instrument was used at a maximum accelerating potential of 25kV in positive mode and was operated in reflectron mode. The sample were prepared by standard dried droplet preparation on stainless steel MALDI targets using α-cyano-4-hydroxycinnamic acid as matrix.
The external calibration of MALDI mass spectra was carried out using singly charged monoisotopic peaks of a mixture of bradykinin 1-7 (m/z=757.400), human angiotensin II (m/z=1046.542), human angiotensin I (m/z=1296.685), substance P (m/z=1347.735), bombesin (m/z=1619.822), renin (m/z=1758.933), ACTH 1-17 (m/z=2093.087) and ACTH 18-39 (m/z=2465.199). To achieve mass accuracy, internal calibration was performed with tryptic peptides coming from autolysis of trypsin, with monoisotopic masses at m/z = 842.510, m/z = 1045.564 and m/z = 2211.105 respectively. Monoisotopic peptide masses were automatically annotated using Flexanalysis 2.0 software.
Peaks are automatically collected with a signal to noise ratio above 4 and a peak quality index greater than 30.

LC-MS/MS

Nano-LC-MS/MS analysis was performed either using a CapLC capillary LC system (Waters), coupled to a hybrid Quadrupole Time-Of-Flight mass spectrometer (Q-TOF II, Waters).
From each sample, 6.4 µL was loaded on a precolumn, before chromatographic separation on a 75 mm id, 150 mm length C18 column (LC Packings C18, 3µm bead size, 10nm mean pore size). The gradient was generated by the CapLC at a flow rate of 200 nL/min. The gradient profile consisted of a linear one from 90% of a water solution acidified by 0.1% HCOOH vol/vol (solution A), to 40% of a solution of CH3CN acidified by 0.1% HCOOH vol/vol (solution B) in 30 min, followed by a second gradient ramp to 75% of B in 1 min. Data acquisition was piloted by MassLynx software V4.0. Calibration was performed using adducts of 0.1% phosphoric acid (Acros, NJ, USA) with a scan range from m/z 50 to 1800. Automatic switching between MS and MS/MS modes was used. The internal parameters of Q-TOF II were set as follows. The electrospray capillary voltage was set to 3.5 kV, the cone voltage set to 35 V, and the source temperature set to 90°C. The MS survey scan was m/z 300–1500 with a scan time of 1 s and an interscan time of 0.1 s. When the peak intensity rose above a threshold of 15 counts/s, tandem mass spectra were acquired. Normalized collision energies for peptide fragmentation were set using the charge-state recognition files for 1+, 2+, and 3+ peptide ions. The scan range for MS/MS acquisition was from m/z 50 to 2000 with a scan time of 1 s and an interscan time of 0.1 s. Fragmentation was performed using argon as the collision gas and with a collision energy profile optimized for various mass ranges of precursor ions.

MS data analysis for PMF
Monoisotopic peptide masses were assigned and used for databases searches using the search engine MASCOT 2.2 (Matrix Science, London, UK) [27]. All proteins present in the Uniprot KB release 12.8 (5503859 protein entries) were used without any pI and Mr restrictions. One missed cleavage per peptide was allowed and the following variable modifications carbamidomethylation for cysteine and oxidation for methionine were taken into account. For Peptide Mass Fingerprint, a mass tolerance of 70 ppm was allowed. A protein was identified by at least 4 peptides and a 20% sequence coverage in addition to the probabilistic score provided by the Mascot algorithm and used at the default threshold.

MS/MS data analysis

Peak lists were automatically created from raw data files using the ProteinLynx Global Server software (version 2.0; Waters) for Q-TOF spectra. The Mascot search algorithm (version 2.2; Matrix Science) was used for searching against the Uniprot KB release 12.8 (5503859 protein entries) The peptide tolerance was typically set to 120 ppm and MS/MS tolerance was set to 0.3 Da in the case of Q-TOF. Only doubly and triply peptides were searched for. A maximum number of one missed cleavage by trypsin was allowed, and carbamidomethylated cysteine and oxidized methionine were set as fixed and variable modifications, respectively. The Mascot score cutoff value for a positive peptide hit was set to 30. Peptides under the cutoff score were Individual checked manually and either interpreted as valid identifications or discarded.

Immunoassay of cytokines
The cytokines were measured in the cell supernatants by ELISA using the OptEIA set for mouse cytokines from Pharmingen according to the procedures recommended by the manufacturer

# 3. Results

Immature dendritic cells secrete extracellular sensors

Activated dendritic cells are known to synthetize many different proteins. In addition to the well known cytokine expression, which is even a test to check for dendritic cell activation with various stimuli [14], [6] [7], recent work has also shown that activated dendritic cells secrete many other proteins, such as proteases or complement proteins [18], [28].
Conversely, immature dendritic cells secrete very few proteins, as shown on figure 1. Among those, some proteins were quite not surprising, as the metalloproteases (e.g. MMP12) [18], which are likely used by the DC to make their way in and out the peripheral tissues. However, we were quite surprised to detect some proteins which can be viewed as extracellular sensors, such as lipocalin 2 and the YM1 protein. Figure 2 shows the mass spectrometry spectra securing these identifications (detailed data are shown on table 1) . Lipocalin 2 is known to bind iron-loaded bacterial siderophores [29], while the YM1 protein is known to bind chitin and its derivatives, as well as heparan sulfates [30, 31]. We reasoned that these proteins could be synthetized to act as extracellular sensors of microbial products, using the same mechanism as soluble CD14 with LPS. To test this hypothesis, we checked the activation of DC by the ligands of these two proteins, namely bacterial siderophores and chitosan.

Chitosan, but not bacterial siderophores, induce partial activation of dendritic cells

To test this hypothesis, immature dendritic cells produced in culture were put in contact either with bacterial siderophores or with chitin degradation products, and the activation was measured by increased membrane expression of MHC class II molecules, stimulation with LPS being used as a positive control. Initial results (not shown) were obtained by co-culturing the DC with chitin powder and a chitinase, thereby mimicking the situation existing in vivo with fungi and protozoa. In order to gain further insight into this action, we also used various commercially-available chitin degradation products (chitosans). The results, shown on figure 3 clearly demonstrate that siderophores do not induce any DC activation, while chitin degradation products could induce this activation. We also tested whether a combination of siderophores and LPS, also offering a better mimick of the situation occurring with Gram-negative bacteria.
This combination of LPS and siderophores, tested in various quantitative ratios, did not induce any modulation of differentiation compared to LPS alone, neither at the level of membrane activation nor at the cytokine levels (data not shown).
Conversely, chitosan of medium molecular mass (greater than 40 kDa) but neither chitin alone nor oligochitosan ( lower than 10 kDa) was able to induce activation at the membrane level. This is in contrast with other polysaccharides [5] for which even oligosaccharides are efficient activators of DC.
To gain further insight into this dendritic cell activation, we tested the membrane expression of the coreceptors, such as CD 80 and CD 86. The results, shown on figure 4, demonstrate that chitosan is able to induce a complete membrane activation of DC, including both the MHC class II and the coreceptors molecules. This remained true whether we used DC produced or not within the presence of or without FLT3-L.

Activation of DC by chitosan uses TLR4

As chitosan is fairly different in its structure from other well-known activators of dendritic cells (for example it is strongly cationic, while most other DC activators are strongly anionic), it was of interest to determine whether the DC activation proceeded via a TLR-dependent mechanism or not, and if yes, which TLR was implicated. From the structure of chitosan, we selected TLR4 as a primary candidate. TLR4 was of special interest, as heparan sulfate, which is the other ligand of YM1, is known to activate DC via a TLR4-dependent mechanism [7]. To test this hypothesis, DC derived from TLR4-deficent mice, were incubated in the presence of chitosan, LPS or pansorbin. The results, shown on figure 5, demonstrate that chitosan induced activation is impaired in TLR4 $^{-/-}$ DC.
.

Activation of DC by chitosan does not induce cytokine production

Classical activators of DC induce both membrane activation and cytokine release [32]. We therefore tested the release of various cytokines upon activation with chitosan. As shown on figure 6, none of the tested cytokines were secreted upon chitosan treatment even when the incubation was prolonged to 48 hours and, whether we used DC produced or not in the presence of with or without FLT3-L (Figure 7) (ajouter aussi sur la figure les temps longs de stimulation, requete du referee). This showed in turn that the activation induced by chitosan cannot be due to bacterial products contaminating the preparation (e.g. LPS or glycolipids), as these products would induce both membrane activation and cytokine release. To cross check whether this absence of detected cytokines could be an immunological artefact induced by the presence of chitosan in the culture medium, we added chitosan to the supernatant of LPS-activated dendritic cells just before the immunoassay for cytokines secretion. The results were similar with and without added chitosan indicating that chitosan did not impair the measurement of the cytokines. Further crosscheck was made by proteomics techniques, which do not use antibodies. The results, shown on figure 8, indicate the presence of TNF spots in LPS-activated cells only, but neither in unstimulated cells nor in chitosan-stimulated cells. This absence of cytokine production turned us to check the functionality of chitosan-activated dendritic cells

Chitosan-activated DC do not stimulate T cells

In order to check the functionality of chitosan-activated DC, we decided to use the mixed lymphocyte reaction. Figure 9 shows that chitosan-activated dendritic cells, as well as immature DC, are unable to stimulate T cells, while LPS-activated dendritic cells stimulate T cells. In the case of double stimulation by both chitosan and LPS, the activated dendritic cells were able to stimulate T cells, showing that the chitosan-stimulated cells are not really anergic cells. At this stage, we decided to test the cytokine profile of such mixed T lymphocyte-DC cultures. This was intended to detect a partial activation of T cells, resulting in cytokine production but not cell proliferation. As shown in figure 10, mixed cultures of T lymphocytes with LPS-activated DC cells produced Il-10, IL-6, interferon gamma and IL-2, showing the typical activation of T cells (IL2 production) with the expected Th1 profile (interferon gamma producing). Oppositely, mixed cultures of T lymphocytes with chitosan-activated T cells produced neither TGF beta, nor IL-10, IL-4, IL-12 and IL-6, and both interferon gamma and IL-2 secretion remained at a very low level, although higher than the one obtained when using unstimulated cells
As DCs are not described to produce IL-2 nor interferon gamma, it can be reasonably inferred that these cytokines arise from the T cells. In this case, the low production of these cytokines in mixed cultures of T cells with chitosan-activated T cells suggests that the T cells are not really anergic under these conditions. However, the lack of proliferation and of CD69 activation suggests that T cells are at most very weakly stimulated by chitosan-activated T cells.

Chitosan does not prevent restimulation of DCs, but reorient it

At this point, it could be argued that chitosan, when present alone on dendritic cells, induces a general block in cytokine production or even a general block in the secretory pathway. To test this hypothesis, we treated dendritic cells first for 24 hours with various doses of chitosan, then with LPS. The results, displayed on figure 11, show that chitosan does not block cytokine production. Surprisingly, we also observed a chitosan-dependent change in the IL10/IL12 balance in the cytokinic profile of stimulated dendritic cells.

## 4. Discussion

A very active area of research aims to understand how dendritic cells are activated or made silent by various signals. Most of the research effort has been devoted to the identification of DC-activators, or danger signals [1, 2], and it is only recently that some insights have been given on how dendritic cells keep immunologically silent when facing apoptotic bodies [33].

With the multiplicity of exogenous and endogenous danger signals identified to date, it appears that dendritic cells are very easily activated, and able to trigger an efficient immune response. This poses in turn the question on how pathogens can still be successful. While acute bacterial infections can be seen as the result of a speed race between pathogen multiplication and immune response, the persistence of chronic infections with slowly-dividing pathogens sets the question on how the pathogens can control the immune response down to a level where the chronic infectious state is reached. While many of such pathogens hide themselves inside the host cells, some (e.g. fungi, some protozoans) do not, and have to face the immune system during their lifetime in the host organism. Perturbing the DC differentiation and/or maturation is a strategy that is obviously followed by some slowly-dividing pathogens, as shown by the examples of M. leprae [34] and P. falciparum [35]. While the mechanism used by M. leprae is still not clear, recent work has shown that red blood cells infected by P. Falciparum are able to prevent both DC activation and the triggering of a Th1 response [35]. It has been shown that this effect can be mediated by the binding of an endogenous glycan, chondroitin sulfate A, to the surface of infected red blood cells [35].

Our results on chitosan stand clearly in the same trend, although they are slightly different. While RBC-bound chondroitin sulfate suppresses both membrane activation and cytokine production by DC, chitosan does induce membrane activation but no cytokine production, resulting in inefficient T-cell priming.

Such uncoupling of membrane activation and cytokine production by DC is not common, however it has been described on a DC cell line [36], suggesting that the two events are not fully linked and can be uncoupled. More recently it has been described that complement receptors are able to achieve such a partial activation, and this has been advocated to result into DC anergy [33]. While our T-cell activation experiments show that this is not real anergy in our case, activation-induced DC death by apoptosis also suggests that such a partial activation will shift the balance toward a less efficient immune system, just by decreasing the number of DC able to present antigens efficiently.

On a more mechanistic point of view, the situation of complex glycans as DC modulators appears quite complex. Lipid-containing glycans (e.g. LPS, Lipoteichoic acids) are well known, fully potent DC activators, following the hydrophobic activator theory [37]. However, lipid-free complex glycans appear to induce very various phenotypes on DC, and this seems to depend on several parameters.

The first one seems to be the type and size of the potential activating complex (e.g. particle of various sizes vs. soluble glucide). For example, large chitin particles have no effect on DC, while medium-size chitin particles have a stimulatory effect [38] and soluble chitosan has either an inhibitory effect at medium size, or no effect at all at low size (our present work). Another example of such an effect is represented by chondroitin sulfate, but in the reverse order. Soluble chondroitin sulfate mediates at least a membrane activation on DC [6] during a long term exposure, but not during a short exposure [7], while the presentation of chondroitin sulfate at the surface of a red blood cell induces DC inhibition [35].

While keeping with soluble glycans only, the length of the sugar also appears to be a critical activation parameter. For example, short chains of hyaluronan have been shown to fully activate DC, while longer chain do not [5]. This has led to the theory in which only fragments of these complex saccharides can act as DC activators, and this has been interpreted as a mechanism for sensing tissue damage [7]. This theory has been extended to several glucan-containing activators, including bacterial products [37] leading to the idea that partial degradation of the cell walls or extracellular matrix complex glucans will increase the reaction of DC. The chitosan example, however, is a clear exception to this rule. Chitosan is produced during the degradation and/or remodeling of the chitin exoskeleton, and our work shows that this degradation product induces an incomplete and therefore inefficient DC reaction.

Taken together with the literature, our results show that subtle differences in the glycans can lead to fairly different DC responses. For example, related anionic glucides, e.g. chondroitin sulfate, hyaluronic acid and heparan sulfate, have quite different effects on DC. It could be argued that this could be due to differences in the membrane receptors and/or soluble co-receptors (analogous to TLR4 + CD14 for LPS) sensing the various glucans. However, the situation seems more complex than that, as heparan sulfate, LMW hyaluronan (full DC activators) and chitosan (partial activator) share a common membrane receptor, TLR4, while heparan sulfate and chitosan share in addition a common soluble binding protein, YM1 [30, 31]. It is not known at the present time whether YM1 is an obligatory co-receptor for these ligands, but it appears quite likely that YM1 by itself does not orient the fate of dendritic cells. As a matter of facts, various YM1 ligands induce different fates for dendritic cells: heparan sulfate leads to a Th1 response[7], simvastatin a Th2 response [39], and chitosan an inefficient response (this work). However, even if the precise role of YM1 in the response of dendritic cells to sugar ligands is unclear so far, the rationale of YM1 as a co-receptor led us to investigate the response of DC to chitosan and to find this peculiar partial activation. Maybe the most intriguing feature of this chitosan response is its sensitivity to the context. While LPS alone induces the classical Th1 response. chitosan alone induces a phenotypic activation without secretion of cytokine, thus inefficient to stimulate T cells effectively, at least at the scale of a immune system response.

When integrating those results at a higher level, it must be kept in mind that we tested CD8+ dendritic cells. The results obtained on a whole organism upon chitosan administration [40] showing high IL10 levels, may result, at least in part, from the response of CD8- dendritic cells [41]. Another fungal cell wall component, zymosan [42], also induces a weak Th2 response of dendritic cells. As zymosan engages TLR2 and chitosan TLR4, which are known to be the major TLR engaged by fungi [43], this may explain, at least in part, the fair resistance of fungi to the immune system.

Acknowledgments.

Anne Marie Laharie is thanked for performing the cytokine immunoassays, Anne Papaioannou for expert assistance in cell culture, Sylvie Luche is thanked for her expert work in proteomics., and Jean Marc Strub for help in the preparation of this manuscript.
David Gray (London) is thanked for the gift of the GM-CSF-producing cell line.
A grant from the Région Rhone Alpes to TR is also acknowledged

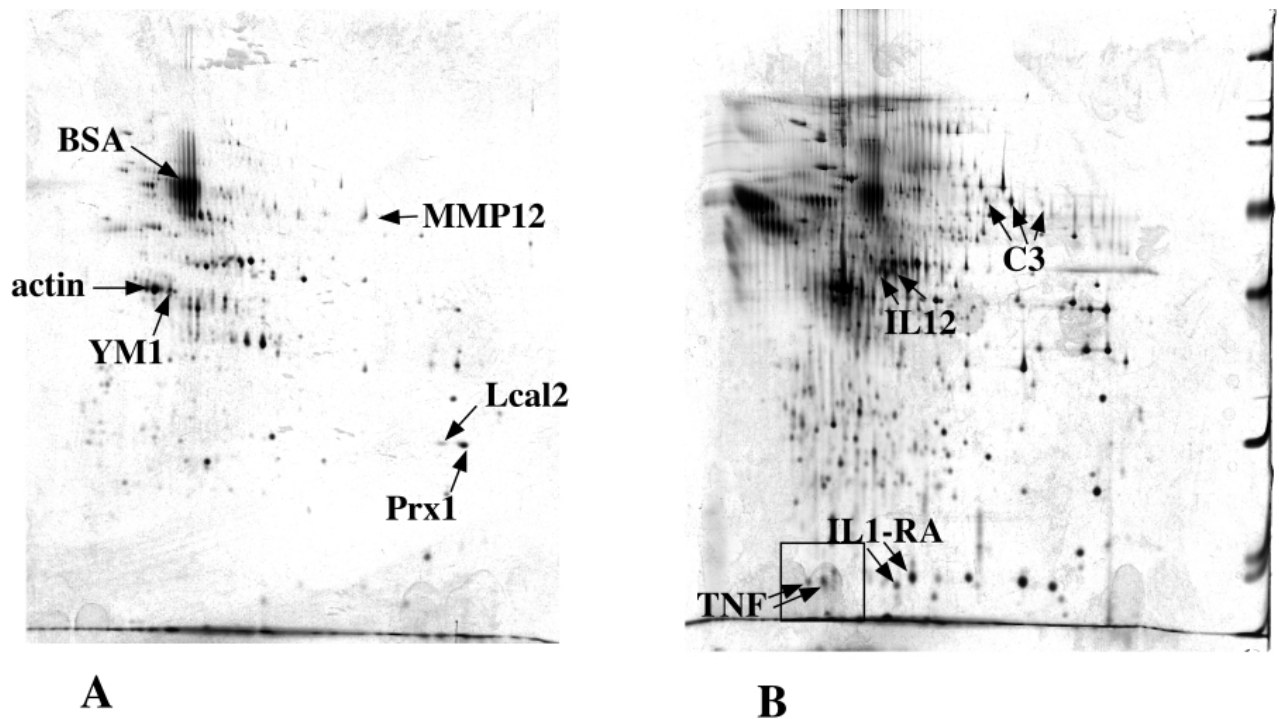

Figure 1: Proteins secreted by dendritic cells
15 ml of medium conditioned by 1 million dendritic cells per ml were concentrated by precipitation. The proteins were then analyzed on two-dimensional gels (3-10 pH gradient in first dimension) and stained with silver 1A: proteins secreted by immature dendritic cells. 1B proteins secreted by dendritic cells activated with 1μg/ml LPS

BSA: bovine serum albumin, arising from the serum used for DC cultivation until the secretion step. Prx1: peroxiredoxin 1 (SwissProt P35700); IL1-RA: interlukin1 receptor antagonist (SP P25085); TNF: TNF alpha (SP P06804); MMP12: macrophage metalloelastase (MMP12) (SP P34960); C3: mouse complement C3 protein, N-terminal fragment (SP P01027); IL12: interleukin 12 beta chain (SP P43432); Lcal2: lipocalin2 (SP P11672); YM1: chitinase3 like protein3 (SP O35744)

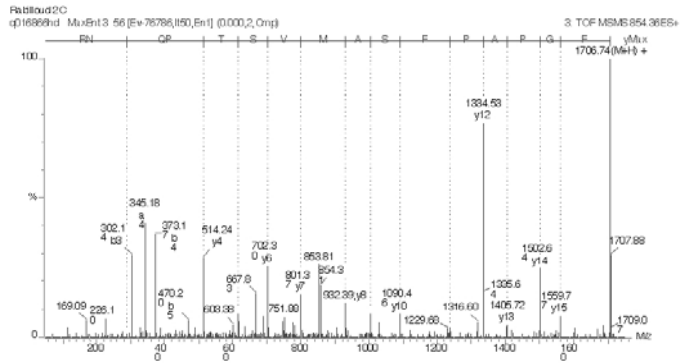

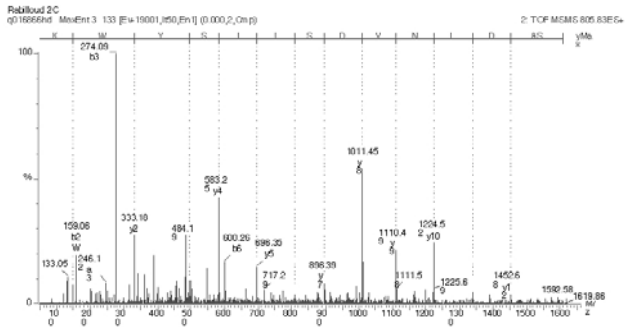

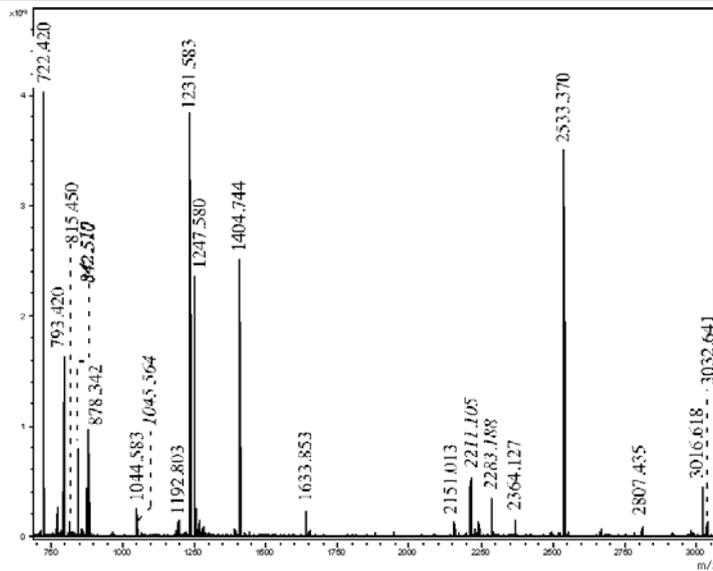

Figure 2: Identification of lipocalin2 and YM1 protein by mass spectrometry
YM1 was identified by MS/MS with 14 different peptides (composite Mascot score 721), and the MS/MS spectra for two of them are shown on panel A. Lipocalin2 was identified by peptide mass fingerprinting (Mascot score 131, sequence coverage 49%, 9 different peptides), and the corresponding spectrum is shown on panel B. On this panel, the italicized masses correspond to trypsin autolysis peptides.

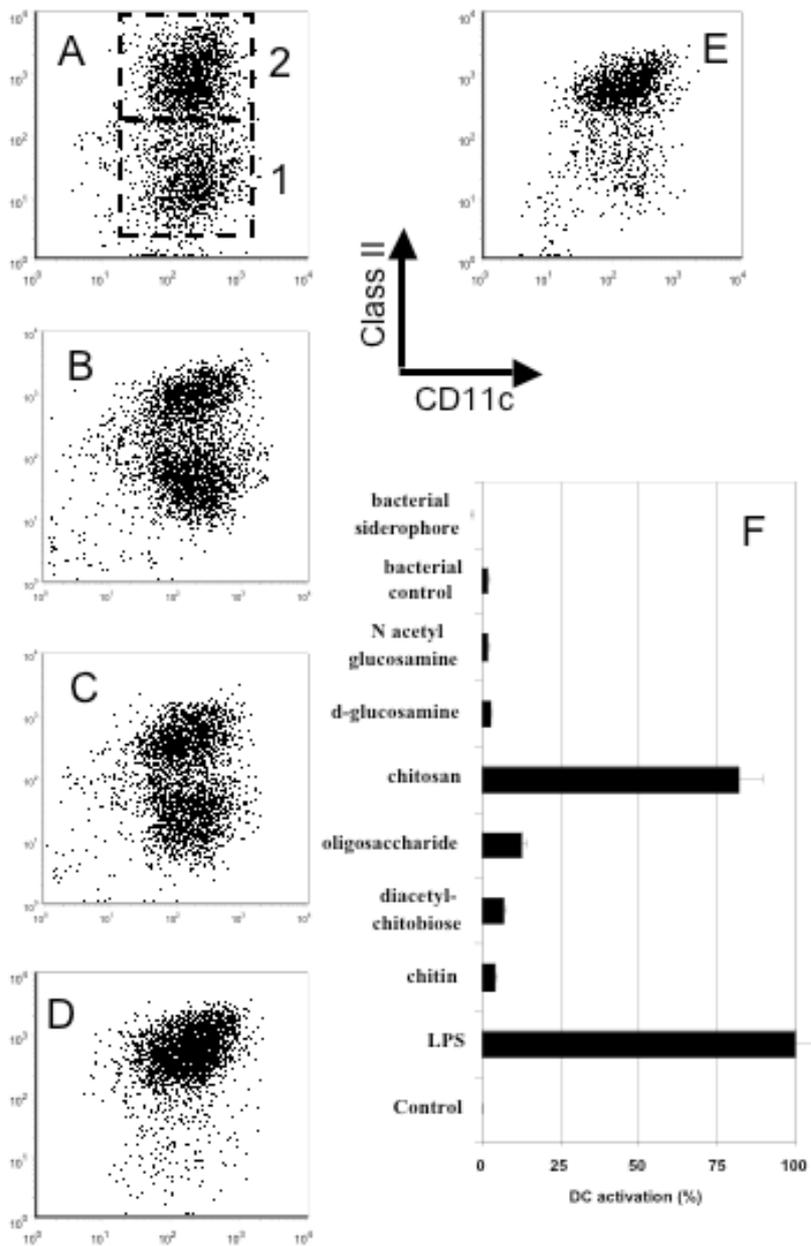

Figure 3: Activation of immature dendritic cells.
DCs were incubated in the culture medium supplemented with the different compounds for 17 hours at 37°C, and then DC activation was assessed by double labelling using fluorescent monoclonal anti-Class II and anti-CD11c molecules. Chitosan (D) and LPS (E) induced activation of DC compared to normal incubation medium (A), to Chitine (B) or to N,N' diacetylchitobiose (C). (F) represents the percentage of DC activation in the presence of various compounds: immature and activated cells were characterized as low and high class II molecules (as shown A box 1 and box 2 respectively), the percentage of activation of DC was calculated from the percentage of the activated cells assuming that the value obtained with the normal incubation medium (A) gave the 0% of activation and the one with LPS (E) gave the 100 of activation.

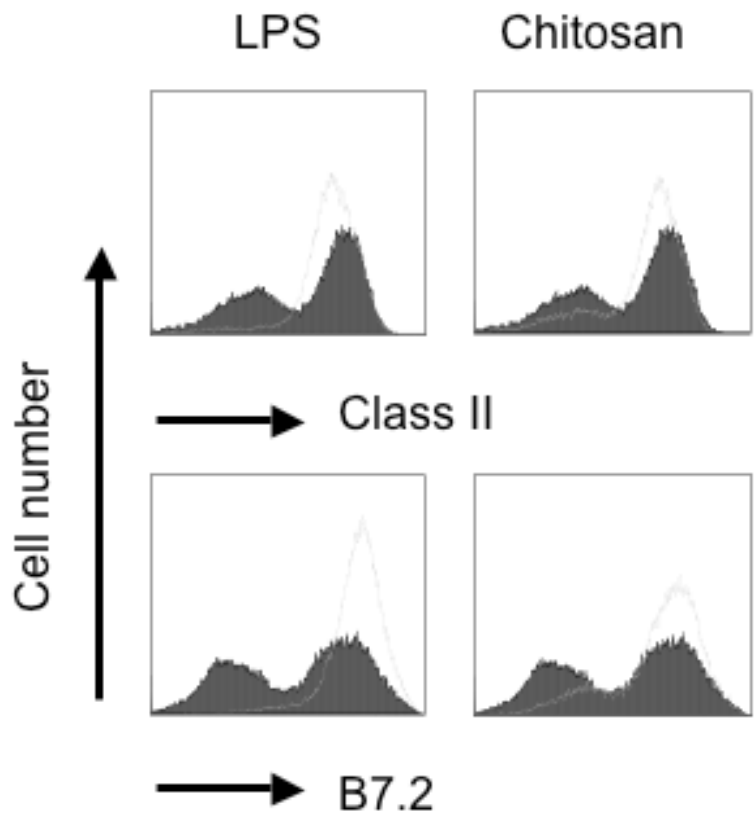

Figure 4: Modification of DC phenotype after incubation in the presence of chitosan.
DCs were incubated in the culture medium supplemented with the different compounds for 17 hours at 37°C, then DC phenotype was assessed by double labelling using fluorescent monoclonal anti-Class II (upper panels) or anti-CD86 (lower panels). The addition of chitosan (grey lines, right panels) increase the expression of Class II and CD86 molecules at the surface of dendritic cells as compared to normal culture conditions (black filled curve), this results are comparable to what is observed in the presence of LPS (left panels).

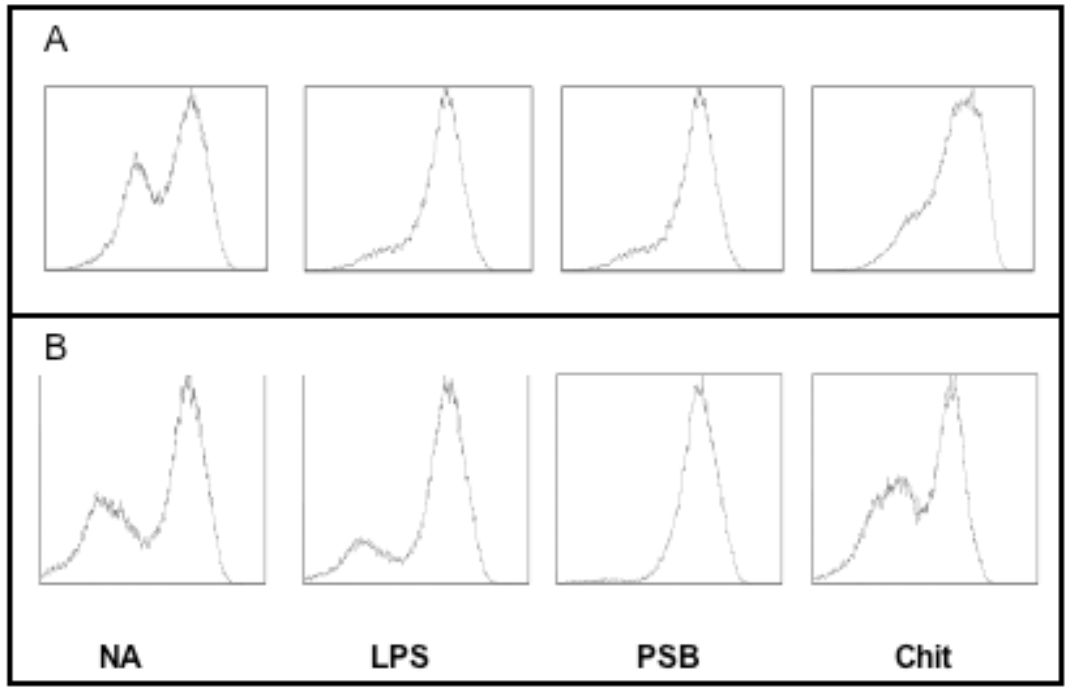

Figure 5: Role of TLR4 in the activation of dendritic cells by chitosan.
Dendritic cells were prepared from progenitors isolated from from wild type C57BL/6 (A) and TLR4-/- (B) mice and incubated with chitosan for 17 hours at 37°C; then expression of Class II molecules was measured to assess the activation. As seen from the disappearance of the low MHC II peak, chitosan (CHIT) increase the expression of class II on the wild type C57BL/6 and not on TLR4-/- DCs. Similar results were obtained in the presence of LPS whereas pansorbin (PSB) activated both type of DC. NA stands for no activator.

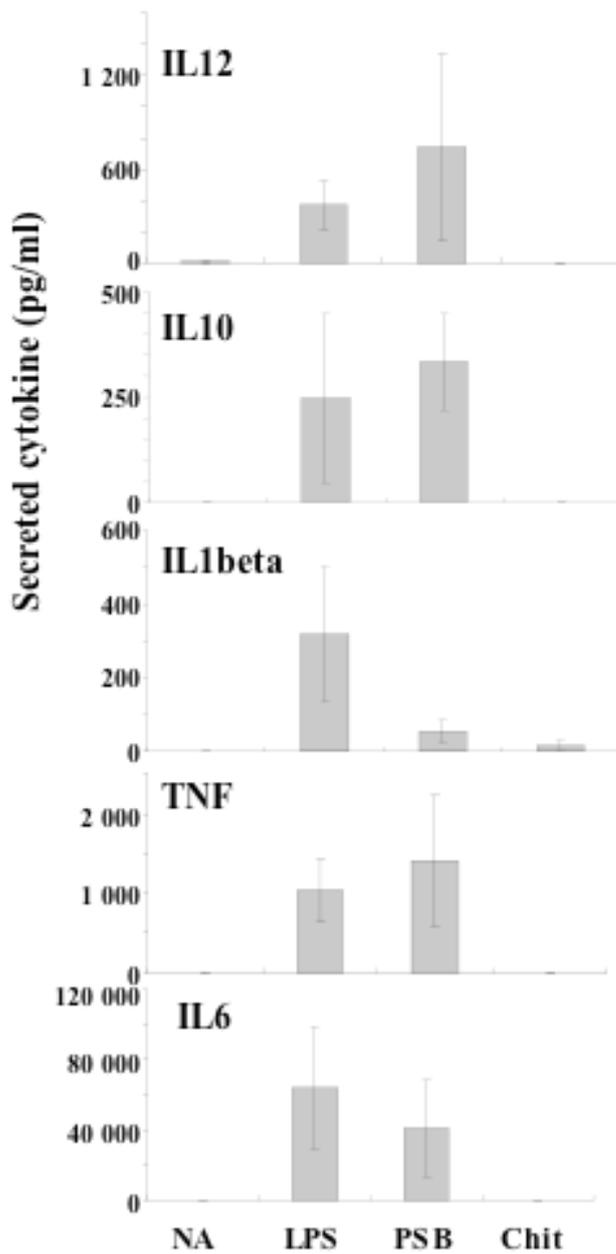

Figure 6: Cytokines secretion of dendritic cells after incubation in the presence of chitosan (chit column) and comparison induced by LPS or pansorbin (PSB).
DCs were incubated with chitosan, LPS or PSB for 17hours at 37°C and the supernatants were measured for their content in cytokines: IL12, IL10, IL1beta, IL6 and TNF alpha and Il6. Similar values were obtained at 48 hours (data not shown).

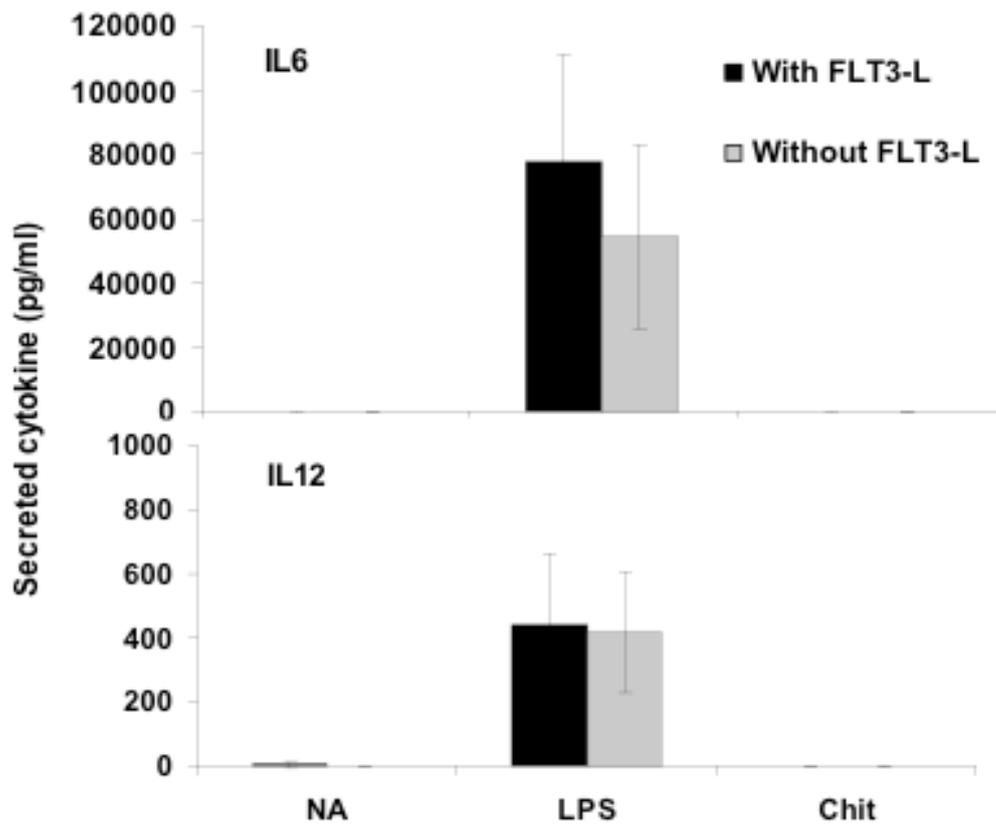

Figure 7:

DCs were cultured in the presence or not of FLT3-L and incubated with chitosan or LPS at 37°C for 24 hours. The supernatants were measured for their content in IL-12 and IL-6 (in pg/ml) by ELISA

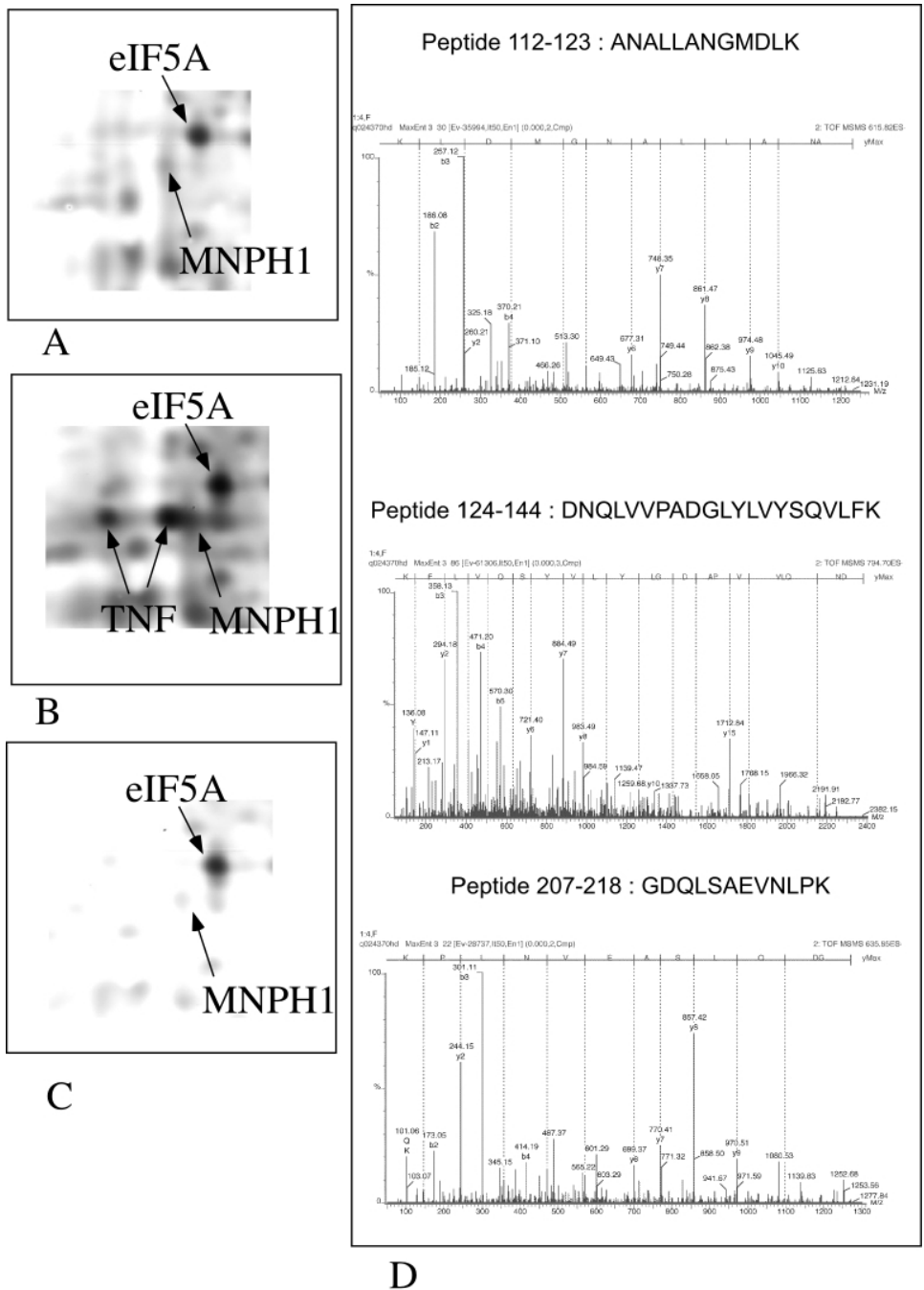

Figure 8: TNF identification by proteomics techniques
25ml of conditioned media were concentrated and analyzed by two-dimensional electrophoresis techniques, as shown on figure 1. Only the inset zone shown on figure 1B is displayed.
7A: immature dendritic cells; 7B dendritic cells activated by 1 µg/ml LPS; 7C: dendritic cells activated by 40µg/ml chitosan. The proteins identified in the inset zone are the following: eIF5A: eukaryotic translation initiation factor 5A (UniProt number P63242); TNF: TNF alpha precursor, C terminal fragment (P06804); MNPH1: mago-nashi protein homolog 1 (Q9CQL1).
Three MS/MS spectra identifying the TNF protein (composite Mascot score 297) are shown in panel D. The data for the other two proteins are shown in supplementary material.

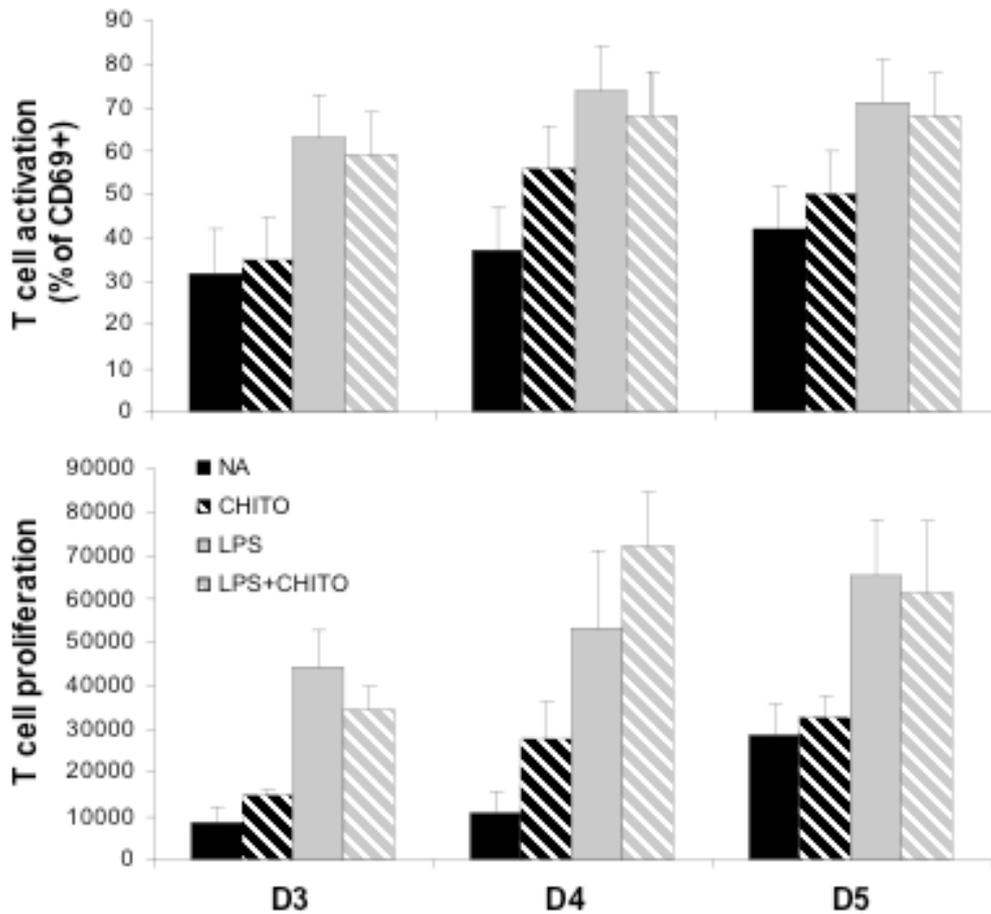

Figure 9: Activation of T lymphocytes by chitosan-activated dendritic cells
DCs were incubated with chitosan or/and LPS for 17hours at 37°C, then DC were washed and treated with mitomycin as described in Material and Methods. DC were then added to T lymphocytes and incubated for the indicated time. T lymphocytes activation was evaluated by measurement of CD69 expression (upper panels) and by their proliferation was measured by the Uptiblue fluorescence (lower panel). Results correspond to 3, 4 and 5 days of incubation of 10 000 DC together with 10 000 T lymphocytes.

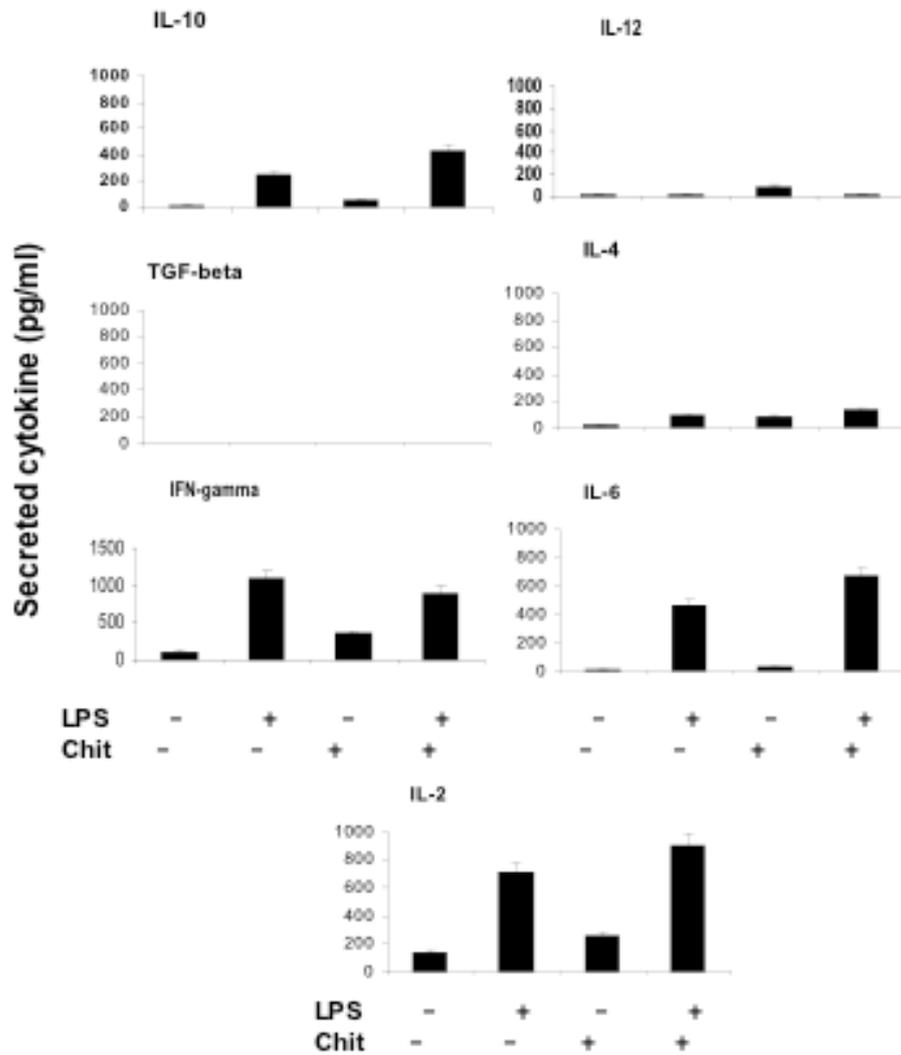

Figure 10: Cytokines secreted during T lymphocytes activation by chitosan-activated dendritic cells
DCs were incubated together with T lymphocytes as described for Figure 8, then the supernatants were collected and their content in cytokines were measured by ELISA, Results correspond to 4 days of incubation of 10 000 DC together with 10 000 T lymphocytes.

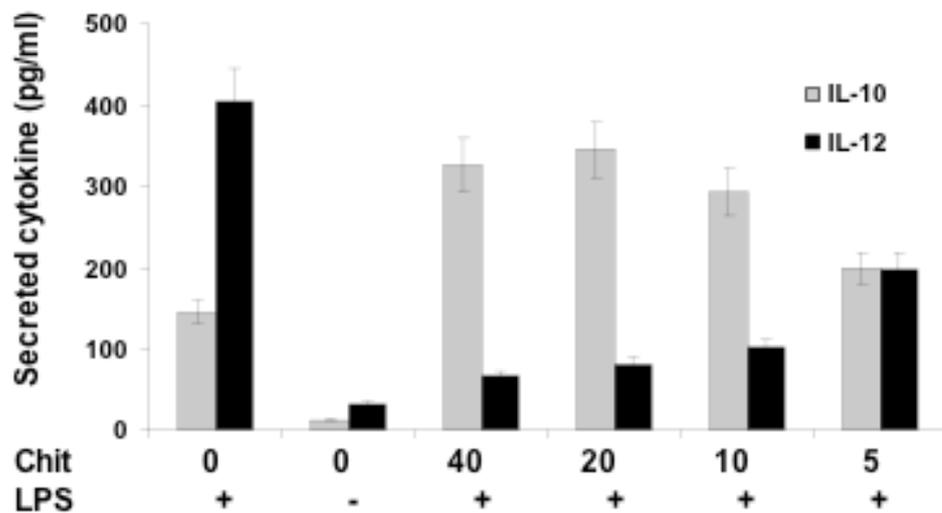

Figure 11: Modification of cytokine production balance by chitosan
DCs were incubated with chitosan at the indicated coincentrations (in µg/ml) at 37°C for 24 hours, then, when indicated, LPS (1 µg/ml) was added to the culture and cells were further incubated at 37°C for 24 hours. The supernatants were collected and the cytokines IL-12 and IL-10 were quantified by ELISA.

Table 1: Identification data
the Uniprot accession number (AC column), name, biochemical parameters and observed peptides are given for the proteins of interest identified. The first two proteins (YM1 and TNF) have been identified by MS/MS, the third one (Lipocalin 2) by pept mass fingerprinting

| AC | Protein | theor. Mass | theor. pI | % cov. | RMS | Score | Nb peptides | Séquences | peptide score. |
|---|---|---|---|---|---|---|---|---|---|
| O35744 | Chitinase-3-like protein 3 precursor -(YM1)-Mus musculus(Mouse) | 44430 | 5.42 | 39 | 82 | 721 | 14 | SPYDIGK | 30 |
| | | | | | | | | HLFSVLVK | 37 |
| | | | | | | | | TLLAIGGWK | 28 |
| | | | | | | | | DYEALNGLK | 41 |
| | | | | | | | | HFPLTSTLK | 41 |
| | | | | | | | | AFEEESVEK | 61 |
| | | | | | | | | QIFIQSVIR | 42 |
| | | | | | | | | KAFEEESVEK | 21 |
| | | | | | | | | DGYTGENSPLYK | 51 |
| | | | | | | | | TGIGAPTISTGPPGK | 67 |
| | | | | | | | | SADLNVDSIISYWK | 69 |
| | | | | | | | | FGPAPFSAMVSTPQNR | 108 |
| | | | | | | | | QYNFDGLNLDWQYPGSR | 63 |
| | | | | | | | | IPELSQSLDYIQVMTYDLHDPK | 62 |
| P06804 | Tumor necrosis factor precursor - Mus musculus (Mouse) | 25879 | 5.01 | 40 | 48 | 297 | 6 | VNLLSAVK | 49 |
| | | | | | | | | ANALLANGMDLK | 55 |
| | | | | | | | | GDQLSAEVNLPK | 52 |
| | | | | | | | | GQGCPDYVLLTHTVSR | 40 |
| | | | | | | | | DNQLVVPADGLYLVYSQVLFK | 54 |
| | | | | | | | | DTPEGAELKPWYEPIYLGGVFQLEK | 47 |
| | | | | | | | | | peptide mass |
| P11672 | Lipocalin2 - Mus musculus (Mouse) | 22861 | 8.96 | 49 | N/A | 131 | 9 (10 matched masses, 6 unmatched masses) | ITLYGR | 722.42 |
| | | | | | | | | TFVPSSR | 793.42 |
| | | | | | | | | ELSPELK | 815.45 |
| | | | | | | | | DQDQGCR (C-CAM) | 878.34 |
| | | | | | | | | TKELSPELK | 1044.58 |
| | | | | | | | | AGQFTLGNMHR | 1231.58 |
| | | | | | | | | WYVVGLAGNAVQK | 1404.74 |
| | | | | | | | | SLGLKDDNIIFSVPTDQCIDN (C-CAM) | 2364.17 |
| | | | | | | | | YPQVQSYNVQVATTDYNQFAMVFFR | 3016.32 |
| | | | | | | | | YPQVQSYNVQVATTDYNQFAMVFFR (M-ox) | 3032.64 |